\documentclass{article}
\usepackage[utf8]{inputenc}
\usepackage{graphicx}
\usepackage{listings}
\usepackage{url}
\usepackage{xcolor}
\usepackage{hyperref}
\usepackage[a4paper]{geometry}
\usepackage[T1]{fontenc}  
\title{CutLang as an Analysis Description Language for Introducing Students to Analyses in Particle Physics}
\author{A. Ad{\i}g\"{u}zel$^{1,4}$, O. \c{C}ak{\i}r$^2$, \"{U}. Kaya$^3$, V. E. \"{O}zcan$^{3,4}$,\\
S. \"{O}zt\"{u}rk$^5$, S. Sekmen$^6$, \.{I}. T\"{u}rk \c{C}ak{\i}r$^7$, G. \"{U}nel$^8$}
\date{
$^1$ \.{I}stanbul University, Physics Dept., \.{I}stanbul, Turkey\\
$^2$ Ankara University, Physics Dept., Ankara, Turkey\\
$^3$ Bo\u{g}azi\c{c}i University, Physics Dept., \.{I}stanbul, Turkey\\
$^4$ Bo\u{g}azi\c{c}i University, Feza G\"{u}rsey Center for Physics and Mathematics, \.{I}stanbul, Turkey\\
$^5$ Tokat Gaziosmanpa\c{s}a University, Physics Dept., Tokat, Turkey\\
$^6$ Kyungpook National University, Physics Dept., Daegu, South Korea\\
$^7$ Giresun University, Dept. of Energy Systems Engineering, Giresun, Turkey\\
$^8$ University of California, Irvine, Physics and Astronomy Dept., USA\\
\vspace{0.7cm}
}

\begin{document}

\maketitle
\begin{abstract}
The fifth edition of the ``Computing Applications in Particle Physics'' school was held on 3-7 February 2020, at \.{I}stanbul University, Turkey. This particular edition focused on the processing of simulated data from the Large Hadron Collider collisions using an Analysis Description Language and its runtime interpreter called CutLang. 24 undergraduate and 6 graduate students were initiated to collider data analysis during the school. After 3 days of lectures and exercises, the students were grouped into teams of 3 or 4 and each team was assigned an analysis publication from ATLAS or CMS experiments. After 1.5 days of independent study, each team was able to reproduce the assigned analysis using CutLang.
\end{abstract}

\section{Introduction}

The utilization of computational tools is an essential requisite for doing research in particle and accelerator physics, yet it is often missing in the curricula of Turkish universities. Based on the experiences of the active researchers working in these fields, the ``Computing Applications in Accelerator and Particle Physics'' school (Turkish acronym being HPFBU) was conceived in 2008~\cite{hpfbu} as a one-intensive-week event for graduate students in Turkey. The aim was to educate new generations of scientists through a direct hands-on attitude, providing them with skills that are readily applicable, and training them by compelling them to perform actual research tasks.

The school has evolved over the years, targeting different sections of the Turkish student pool and updating its content to keep in accordance with the latest developments in the field. Its fifth iteration, focused completely on data analysis of the Large Hadron Collider (LHC) data, was held between 3-7 February 2020, and targeted mostly undergraduate students, along with a few embedded graduate students. To ease their task, especially for the pre-literate ones in terms of computer programming languages, a recently developed approach, called an analysis description language was introduced.

In high energy physics, collider data analyses are performed using analysis software frameworks that are based on general purpose languages like C++ or Python.  In these frameworks, physics content and technical operations are intertwined and handled together, which makes it very difficult for beginning students to be involved in the research process.  In the last couple of years, the particle physics community has been investigating alternative ways that would allow for a more direct interaction with data through decoupling physics information from purely technical tasks. One effort in that direction introduced the concept of analysis description language, which is a domain-specific and declarative language designed to describe the physics content of a collider analysis in a standard and unambiguous way~\cite{Brooijmans:2016vro, Sekmen:2020vph, adlweb}. This idea was concretely realized in the recently developed interpreted language CutLang~\cite{Sekmen:2018ehb, Unel:2019reo, cutlanggithub}.

CutLang uses a plain text Analysis Description Language file (ADL file),  containing blocks with a keyword-value structure. The blocks make clear the separation of analysis components such as object definitions, variable definitions, and event selections while the keywords specify analysis concepts and operations. The syntax includes mathematical and logical operations, comparison and optimization operators, reducers, four-vector algebra and common HEP-specific functions (e.g. $\Delta\phi$, $\Delta R$, etc.). ADL files can refer to self-contained functions encapsulating variables with complex algorithms (e.g. stransverse mass $M_{T2}$~\cite{RMT2}, aplanarity, etc.) or non-analytic variables (e.g. efficiency tables, machine learning discriminators, etc.). 

When the description of the physics algorithm is completely separated from the software frameworks, any framework capable of interpreting the language that describes physics can be used for performing an analysis on events. CutLang achieves this goal by using a runtime interpreter~\cite{Sekmen:2018ehb, Unel:2019reo, cutlanggithub}, capable of operating directly on events without the need for compilation. Not having the necessity to write code or compile it, combined with the simple, human-readable nature of the domain-specific language syntax makes it a practical construct for quickly performing phenomenological and educational analyses as the ones in the school.

\section{School Organization and Program}
\subsection{Student Selection}
Although the previous iterations of the school had targeted graduate students only, this year, partially motivated by the premise of ADL lowering the entry barrier to the field, most participants were selected from amongst undergraduate students majoring in physics or engineering at various departments throughout Turkey. 
Following a call mostly through social media, about 132 applications were received in about two months. 24 undergraduate, 3 MSc and 3 PhD students were accepted after consideration of their GPAs, resumes, reference letters and their scores on a simple web-based quiz focusing on the very basics of particle physics. About one third of the students were affiliated with some engineering department. No prior computer, programming or operating system knowledge requirements were imposed during the selection process.

\subsection{Logistics}
The event took place in the main campus of \.{I}stanbul University, and was hosted by the Astronomy and Space Sciences Department. While the past iterations of HPFBU were held at various cities throughout Turkey, the more focused nature of this iteration played a key role in the choice of \.{I}stanbul as the location, since most of the domestic instructors with past ADL experience are affiliated with universities within the city. The venue was selected taking into account its central location, its ease of access through public transportation, and the participation of \.{I}stanbul University to both ATLAS and CMS experiments, with their contributions to ATLAS data analyses performed using ADLs.

The student dormitories of the university were allocated to participating students (including those who normally reside nearby) to allow the discussions and peer-to-peer education to continue up to as-late-as-possible hours. The lectures were both recorded and broadcast live on the internet. (Lecture slides and recordings have since been made publicly available on the school's Indico webpage~\cite{pfbuindico}.) The computing exercises were conducted in the allocated computer room with 15 desktop machines. Students who preferred to use their own laptop computers were requested to have downloaded a virtual computer image and installed the Virtual Box host application~\cite{virtualbox} before attending the exercise sessions.

\subsection{Lecture Contents}
The lectures start with the fundamentals of elementary particles and introduce the necessary physics and mathematics background for understanding the data analysis in particle physics. The basics of the relevant computer software (text editors, command shell, ROOT) are also presented, followed by some hands-on training on their utilization.

Recently, utilization of a domain-specific analysis description language as an alternative to general purpose languages (e.g. C++ or Python) is becoming popular as they allow faster and more efficient implementation of analyses, and ease in translating one analysis through different stages in its life cycle, such as phenomenological feasibility, tests with detector models, and actual analysis with data. So the lectures have been set up to demonstrate object selection, object reconstruction and event selection through examples with CutLang, the analysis description language and runtime interpreter.

The core goal of the school is to bring students to a level that allows them to perform rudimentary forms of actual analyses of Monte Carlo pseudo-data and real data from the LHC experiments (obtained through the open data initiative). Towards that goal, a couple of lectures broadly cover the statistical techniques as well.

\begin{figure}
    \centering
    \includegraphics[scale=0.2]{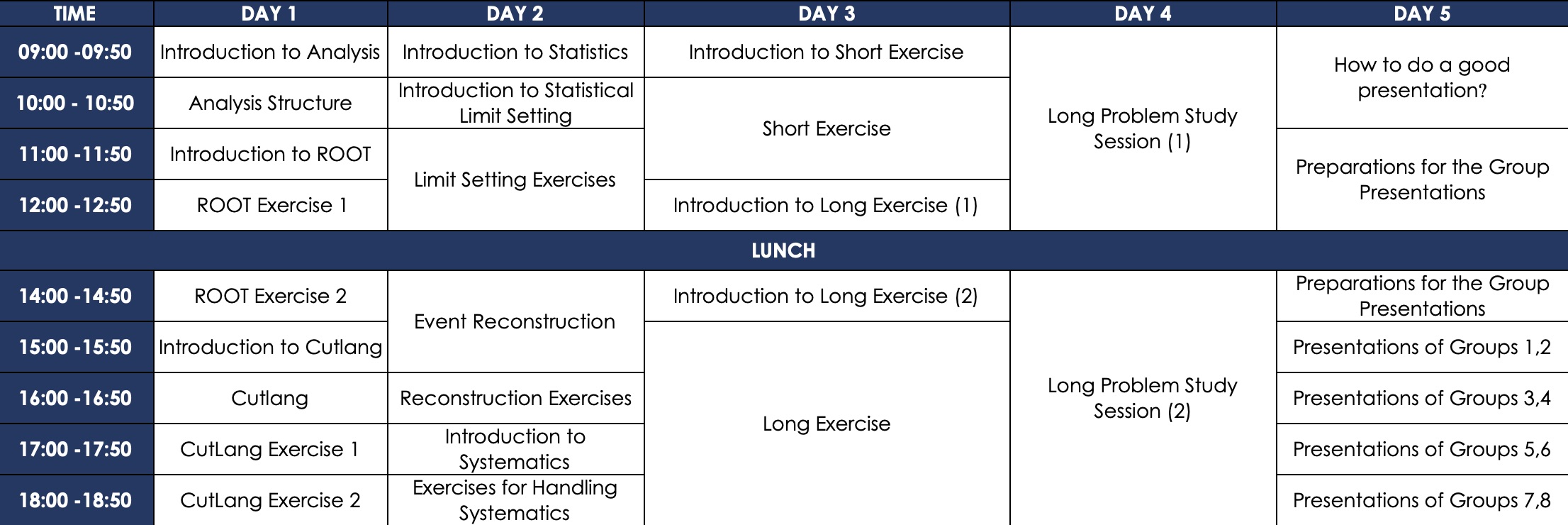}
    \caption{PFBU School Program}
    \label{fig:my_label}
\end{figure}

 \subsubsection{Introduction to Analysis}
 This lecture starts from the basics of elementary particles and interactions. The most relevant concepts for experimental research are presented, such as the process cross sections and branching fractions. Particle detectors are then introduced as a means of observing known particles and searching for new particles. The concept of a particle jet is discussed. For the students to appreciate what happens at a collision, the kinematic variables and hard scattering are discussed in some detail. The importance of reconstructing invariant mass as a discriminant is presented with examples from $Z$ and $W$ bosons, which motivates the teaching of concepts such as transverse momentum, pseudorapidity and missing energy.
 
 \subsubsection{Analysis Structure}
 This lecture is envisaged as a practical approach to HEP analysis concepts. It starts with a description of quantum field theory and Feynman diagrams, showing in particular the vertices in quantum electrodynamics, quantum chromodynamics and weak interactions. No attempt is made to describe how to do matrix element computations, but conservation laws are covered. What is meant by a collision event is defined, and then the main steps of data analysis are presented: 1) defining a signal and selecting an appropriate search channel, 2) defining a trigger for signal events (the trigger's function is introduced at this point), 3) reconstruction of new objects and eventually signal events, 4) estimation of the contribution from background events (kinds of backgrounds are introduced at this stage), 5) statistical analysis. Finally, the role of systematic uncertainties in proper interpretation of the experimental results is also mentioned. Details of all these concepts are left to the following lectures.
 
 \subsubsection{Introduction to ROOT}
 Taking into account the need for a minimum command line know-how to edit a text file, to look at a histogram, etc., this lecture starts with an introduction to Linux and terminal commands. The rest of the lecture introduces ROOT~\cite{root}, its command line interface and simple calculations through that interface, its basic graphing faculties focusing mainly on histograms and functions. These concepts are solidified through exercises executed with simple macros. Second part of the lecture brings more advanced concepts like error representation, histogram operations (normalization, scaling, stacking, etc.) and finally ntuple preparation, recording into ROOT files and reading back.
 
 \subsubsection{Introduction to CutLang: summary}
 This lecture first describes to the students what an analysis description language is and the motivation to use one. The basic syntax rules for describing an analysis in CutLang~\cite{Sekmen:2018ehb, Unel:2019reo, cutlanggithub} are introduced, as well as the means of accessing and utilizing its runtime interpreter. This interpreter enables CutLang to operate directly on events without the need for any code compilation, which allows the students to directly observe the outcomes of operations they perform on the data. Moreover, while the interpreter itself is implemented in C++ and the underlying technology involves ROOT classes, automatically generated dictionaries and grammar rules based on Unix tools Lex and Yacc~\cite{lexyacc}, it isolates the users from such technical details, hence allowing the students to simply focus on the analysis tasks. The typical output of an analysis in CutLang is a file containing cutflows, ROOT histograms defined by the users (to be used for subsequent statistical analysis), and optionally, the surviving events.  Cutflows are also provided as a text output. The lecture covers all the basic steps of data analysis using examples and exercises executed using CutLang.

 \subsubsection{Introduction to Probability and Statistics}
 Basic probability distributions (starting with Poisson statistics) are introduced via real life examples such as proton decay, integrated with toy Monte Carlo experiments implemented in rudimentary ROOT macros. Bayesian and frequentist statistical methods are discussed within the context of hypothesis testing and limit setting. False positives and negatives are discussed together with $\chi^2$ distribution for decision making within the context of particle physics analysis. Particle identification and signal exclusion are presented as relevant examples, with some discussion of topics like the look-elsewhere effect and the interpretation of exclusion plots. Finally a simple limit setting software, TLimit from ROOT, is shown with a working example.
 
 \subsubsection{Particle Reconstruction}
 The first part of the lecture starts with the details of ATLAS and CMS detectors to focus on particle isolation, identification and reconstruction. Methods such as Particle Flow are discussed, as well as particle tracking and vertex identification. The performance of various LHC tracking detectors are compared and discussed. Muon and $b$-jet identification methods using track fitting and $\chi^2$ minimisation are presented. Calorimeters are introduced within the context of electron, photon and tau reconstruction and identification.
 
 The second part of this lecture is solely dedicated to jet reconstruction. It aims to familiarize the student with strong interactions, hadronic collisions and jet structure. To that end, it also reviews parton distribution functions (PDFs) and hard scattering. Details of jet reconstruction algorithms like cone, $k_T$ and anti-$k_T$ are presented. Finally sub-jets, jet-charge and jet energy calibration topics are discussed.
 
 \subsubsection{Introduction to Systematic Uncertainties}
 After introducing the concept of uncertainties, accuracy and precision, the main uncertainty sources are discussed: detector simulation and reconstruction, event recording and trigger and HEP theory. Details of the experimental uncertainties are elaborated: detector calibration, efficiency, and resolution. Their impact is discussed from both physics and statistics point of views. The concept of nuisance parameter is introduced for simple application in frequentist and Bayesian statistics. PDF uncertainties are given as an example to theoretical uncertainties. Simple exercises are performed using ROOT's TLimit function to enforce learning. 

 \subsubsection{Beyond SM Review}
 After a brief review of the shortcomings of the Standard Model (SM), its four pillars are presented: the particle content, the forces, the symmetry group and the dimensions of space-time. Possible modifications for each of these four items are discussed as well as the possible solutions to the Standard Model's shortcomings. For each case, searches from the LHC experiments and the limits for various models together with important model parameters are shown. Finally, future accelerators and research plans are summarized to motivate the students.

 \subsubsection{Presentation Tips}
 Some tips on how to present results, how to display data, and how to prepare clear slides are discussed. Examples from ugly and cumbersome slides are given to show what not to do. Practical advice is given on how to keep analysis presentations concise and how to fit key messages within limited duration that are commonly allocated to presentations in particle physics events.

\subsection{Short Exercises} \label{sec:shortex}
The students are offered a set of short but complete exercises to get used to ROOT file manipulation, CutLang ADL file writing, CutLang utilization and histogram interpretation. Here, we present one illustrative, medium-level short example for writing a CutLang analysis. The rest of the examples are presented in Appendix~\ref{appendix:B}.

The example features the process $pp \rightarrow WW \rightarrow \mu \nu jj$ . The final state consists of a muon, two jets and MET. Initially both leptonically and hadronically decaying $W$ bosons are defined. The muons and jets are denoted as MUO and JET in CutLang, respectively.  The neutrino in the final state is assumed to have been measured as the missing transverse energy (MET), which, combined with the charged lepton, is used to reconstruct the $W$-boson candidate. In this case, the Lorentz vector from MET, denoted as METLV in CutLang, is defined as ($MET, MET_x, MET_y, 0 $). The muon and MET are combined to reconstruct a leptonically decaying $W$ boson, which is defined as WLreco.  Similarly, two jets are combined to form a hadronically decaying $W$ boson, which is defined as WHreco.  The attributes of objects such as MUO, JET, METLV, WLreco and WHreco, e.g. transverse momentum, pseudorapidity or mass are denoted with the notation {}Pt, {}Eta, {}m or alternatively Pt(), Eta(), m(). The particles in each object collection are initially sorted in increasing order of transverse momentum; therefore the particle with the highest transverse momentum is referred to as [0]  or \_0.  Number of particles in a given collection are counted with Size().  All object, attribute and function names in CutLang are case-insensitive.  

The event selection in the $pp \rightarrow WW \rightarrow \mu \nu jj$ example starts with a requirement of at least one muon and sufficiently large MET ($>20$\,GeV). The events are also required to have at least two jets. Finally, the $W$ boson candidate masses are obtained from the highest momentum physics objects and are plotted for both leptonic and hadronic cases separately. The ADL file for this exercise is as follows: 

\begin{lstlisting}
define WLreco : MUO[0] METLV[0]
define WHreco : JET[0] JET[1]

region   test 
  select ALL # Used for counting all events
  select Size(MUO) >= 1
  select MET > 20
  select Size(JET) >=2 
  histo  h1mWL,  "W_L candidate mass (GeV)", 100, 0, 200, {WLreco}m
  histo  h2mWH,  "W_H candidate mass (GeV)", 100, 0, 200, {WHreco}m
\end{lstlisting}

\subsection{Team Analyses}
During the last two days of the event, the students were grouped into teams of 3-4 students. Each group was assigned a particular ATLAS or CMS analysis and was expected to skim through the relevant publication and then write their own CutLang ADL files. The assigned analyses are listed below:

\begin{enumerate}
    \item ATLAS Collaboration, ``Search for new resonances in mass distributions of jet pairs using 139\,fb$^{-1}$ of $pp$ collisions at $\sqrt{s}=13$\,TeV with the ATLAS detector'', JHEP 03 (2020) 145, \href{https://arxiv.org/abs/1910.08447}{arXiv:1910.08447}.
    \item ATLAS Collaboration, ``Search for high-mass dilepton resonances using 139 fb$^{-1}$ of pp collision data at $\sqrt{s}=13$\,TeV with the ATLAS detector'', Phys. Lett. B 796 (2019) 68, \href{https://arxiv.org/abs/1903.06248v3}{arXiv:1903.06248v3}.
    \item ATLAS Collaboration, ``Search for a heavy charged boson in events with a charged lepton and missing transverse momentum from pp collisions at $\sqrt{s}=13$\,TeV with the ATLAS detector'', Phys. Rev. D 100, 052013 (2019),  \href{https://arxiv.org/abs/1906.05609v2}{arXiv:1906.05609v2}.
    \item ATLAS Collaboration, ``Search for resonances decay into into top-quark pairs using fully hadronic decays in pp collisions with ATLAS at $\sqrt{s}=13$\,TeV'', JHEP 1301 (2013) 116,  \href{https://arxiv.org/abs/1211.2202v3}{arXiv:1211.2202v3}.
    \item CMS Collaboration, ``Measurement of the top quark pair production cross section at 13\,TeV with the CMS detector'', Phys. Rev. Lett. 116, 052002 (2016),  CMS-TOP-15-003,  \href{https://arxiv.org/abs/1510.05302}{arXiv:1510.05302}.
    \item CMS Collaboration, ``Search for supersymmetry in multijet events with missing transverse momentum in proton-proton collisions at 13\,TeV'', Phys. Rev. D 96, 032003 (2017), CMS-SUS-16-033, \href{https://arxiv.org/abs/1704.07781}{arXiv:1704.07781}.
    \item CMS Collaboration, ``Search for supersymmetry in events with one lepton and multiple jets exploiting the angular correlation between the lepton and the missing transverse momentum in proton-proton collisions at $\sqrt{s}=13$\,TeV'', Phys. Lett. B 780 (2018) 384, CMS-SUS-16-042,  \href{https://arxiv.org/abs/1709.09814}{arXiv:1709.09814}.
    \item CMS Collaboration, ``Search for new physics in events with two soft oppositely charged leptons and missing transverse momentum in proton-proton collisions at $\sqrt{s}=13$\,TeV'', Phys. Lett. B 782 (2018) 440, CMS-SUS-16-048, \href{https://arxiv.org/abs/1801.01846}{arXiv:1801.01846}.
\end{enumerate}

\subsubsection{Generation of the Simulated Data}

Monte Carlo event samples used for performing the analyses in this school were either produced by the instructors or taken from the CERN LHC Open Data portal~\cite{opendata}.  Signal samples used in the SUSY exercises were generated using Pythia8~\cite{pythia8}.  Signal models, consisting of several gluino, top squark or electroweak gaugino pair production benchmarks were defined by SUSY Les Houches Accord files~\cite{slha}. Signal and SM background samples in dijet and dilepton resonances analyzes were also produced using Pythia8. The output from Pythia8 was recorded using the HepMC format for an easy interface with detector simulation~\cite{hepmc}. Detector response for all signal samples was modelled with the Delphes detector simulator using the default ATLAS or CMS detector cards~\cite{delphes}. Background samples including $tt+$jets, $W+$jets, etc. were directly taken from the ATLAS open data.

\subsection{Student Presentations}
The students, after their independent analysis training, were expected to work in teams and prepare short presentations. The importance for a HEP career of being able to defend results in front of one's peers was emphasized by the instructors, and so each member of every team was required to present some aspect of their team's analysis, even if that meant discussing 2 or 3 slides. Out of the 8 teams, while none were able to extract limits (since not all backgrounds were available), all were able to obtain various object histograms, 5 were able to produce the final discriminant histograms, and 2 were even able to do fits to those histograms with ROOT. The student presentations can be accessed from the indico page of the school~\cite{pfbuindico}.

\subsubsection{An Example Analysis from the Team Exercises}

In order to illustrate the description of a comprehensive event selection with CutLang, we present below the ADL file for one example case out of the 8 analyses studied in the school: the CMS supersymmetry analysis ``Search for supersymmetry in events with one lepton and multiple jets exploiting the angular correlation between the lepton and the missing transverse momentum in proton-proton collisions at $\sqrt{s}=13$\,TeV'', Phys. Lett. B 780 (2018) 384, CMS-SUS-16-042,  \href{https://arxiv.org/abs/1709.09814}{arXiv:1709.09814}.  The ADL files for implementations of all the analyses can be accessed from the school's indico page~\cite{pfbuindico}.

The description starts by applying kinematic selections to input objects such as electrons, muons, jets in {\tt object} blocks to obtain new electron, muon, jet and and $b$-jet collections.  Additionally, electrons and muons are grouped into lepton collections.  Electrons, muons and leptons to be used for event selection and event veto are separately defined. Next, a leptonically decaying $W$ boson is reconstructed and event variables, such as LepT, the sum of first lepton $p_T$ and MET, dphilW, the angular separation between the first lepton and MET, and HT, the hadronic transverse momentum are defined.  This is followed by the event selection and the definition and filling of histograms in the {\tt region} blocks.  Event selection is done in two steps, first by defining a common preselection region, followed by defining two main signal regions, with and without $b$-jets, derived from the preselection region.

\begin{lstlisting}

# Object definitions
object goodEle : ELE
  select abs({ELE}Eta) < 1.44
  select  Pt(ELE) > 25

object goodMuo : MUO
  select Pt(MUO) > 25
  select abs({MUO}Eta) < 2.1

object goodLep : Union(goodEle , goodMuo)

object goodJet : JET
  select Pt(JET)       > 30
  select abs({JET}Eta) < 2.4
  select dR(JET,goodLep) > 0.4

object vetoEle : ELE
  select Pt(ELE) > 10 AND Pt(ELE) < 25

object vetoMuo : MUO
  select Pt(MUO) > 10 AND Pt(MUO) < 25

object vetoLep : Union(vetoEle , vetoMuo)

object bjet : goodJet
 select bTag(goodJet) == 1

# Variable definitions
define Wlep = goodLep[0] METLV[0]
define LepT = pT(goodLep[0]) + MET
define dphilW = dPhi(goodLep[0], Wlep)
define HT = fHT(jets)

# Event selection and histograms
region preselection
  select ALL
  select Size(goodEle) >= 0
  select Size(goodMuo) >= 0
  select Size(goodLep) == 1
  select Size(vetoEle) >= 0
  select Size(vetoMuo) >= 0
  select Size(vetoLep) < 1
  select Size(goodJet) >= 5
  select Pt(goodJet[0]) > 80
  select Pt(goodJet[1]) > 80
  select HT > 500 
  select LepT > 250
  select dphilW > 0.5
  histo hlep1pt, "1st lepton pT (GeV)", 50, 20, 220, Pt(goodLep[0]) 
  histo hjet1pt, "1st jet pT (GeV)", 50, 80, 1080, Pt(goodJet[0])
  histo hHT, "HT (GeV)", 100, 500, 1500, HT
  histo hdphilW, "dphi(lepton, W)", 50, 0, 3.14, dphilW
  histo hLepT, "LT (GeV)", 50, 250, 1000, LepT

region nob
  preselection
  select Size(bjet) == 0

region multib
  preselection
  select Size(bjet) > 0
  select Size(goodJet) >= 6

\end{lstlisting}

The actual CMS analysis has a further division of the signal regions, which was not implemented by the students, and is not shown here.  A more complete ADL file for the analysis as well as for 15 other LHC analyses can be found in a growing github repository in~\cite{adllhcanalyses}.

\section{Conclusions}
On the whole, the school has reached its goal of engaging undergraduate students in particle physics data analysis. While we lack any standardized means to measure competence in particle physics, our oral evaluations of the students at the conclusion of their final presentations have provided us very encouraging data regarding the accessibility of HEP data analysis by junior and senior-level university students.

For many concepts or jargon that students are expected to have heard before, either in popular physics literature (e.g. top quark) or in other physics courses (e.g.  transverse momentum, cross-section), the descriptions provided, as well as questions asked by the students, provided cues that were indicative of the depth of their understanding. Their presentations included descriptions of detectors like ATLAS and CMS, as well as discussions of final state objects such as electrons, muons and jets. Their attention to particular details, for instance why a certain object needs to be identified with specific selection criteria, or what part of a detector is relevant to the analysis at hand, also provided valuable clues. 

The inferences from our oral evaluation were supported by the data obtained from two separate questionnaires, one filled anonymously by all the students right after the event and another conducted through a web form 9 months after. Based on both Likert-scale questions and open-form comments, it was seen that the school generally met student expectations. The students found the content, theoretical and application-wise, appropriate for their level, were satisfied about the competence and communication skills of the educators, and expressed their enthusiasm for the field. Three out of 23 responders to web questionnaire revealed that they had since joined a research group where they are able to use skills acquired through the school.

The web questionnaire also allowed a test of the level of retention for some of the key knowledge imparted at the school. For example, three quarters of the responding students indicated that they were comfortable with particle jets, and about half were comfortable with pseudorapidity and branching fraction. 87\% agreed that they would be able to prepare a simple Cutlang ADL file either by themselves or with their peers, using the school materials.

The only aspects of the school that the students expressed a need for some improvement were the limited duration and the logistics of the computing classroom. Regardless, when asked whether they would recommend the school to other students in possible future iterations, the average Likert score was significantly positive: 4.8/5.0 (std. dev. 0.8).

We are convinced that introductory level HEP data analysis, including the associated topics of MC techniques and statistical methods, appears to be an appropriate topic for undergraduates if the computer and programming technicalities can be surmounted. Performing such an analysis could also be integrated into the syllabus of a regular particle physics course, as it offers a concrete task that is sorely missing in such courses. The utilization of an Analysis Description Language greatly lowers the barrier of entry, making actual data analysis accessible even to students with no prior programming experience.

Particle physics community, organized through networks like the International Particle Physics Outreach Group, is known for its collaborative efforts to bring basic science to the young individuals through events like the International Masterclasses~\cite{intmasterclass}. Yet, between such events that target high-school students and the on-job training of post-graduate students, there is almost a vacuum in the teaching of experimental particle physics. We have observed that the use of a domain-specific analysis description language such as CutLang provides a bridge in between. We encourage our colleagues, not only to integrate the analysis description language approach and CutLang into their research, but also to engage undergraduate students with schools similar to ours and share their experience.

\section{Funding and Acknowledgements}
This school has been supported by Bo\u{g}azi\c{c}i University Foundation (B\"{U}VAK) Prof. Engin Ar{\i}k Funds and \.{I}stanbul University rectorate. We would like to thank all the students for their enthusiasm and hard work. We also thank Sabine Kraml for her suggestion on presenting the results of the school in such a document and for useful discussions.  SS is partially supported by the National Research Foundation of Korea (NRF), funded by the Ministry of Science \& ICT under contract NRF-2008-00460. GU would like to dedicate this paper to the memory of his late physics teacher, Frère Pierre Caporal who had introduced him to the world of particles.

\appendix
\section{Photos from the School}
\label{appendix:A}
\renewcommand{\thesubsection}{A.\arabic{subsection}}

Here we present some photos from the lectures, exercise sessions and presentations from the school.  Additional group photos with students and instructors are also included. The authors have asked for consent from the school participants for publication of school photographs in this document through an online survey.  All respondents have consented to the publication of these photographs, except one, whose face has been pixellated.

\begin{figure}[!htb]
    \centering
    \includegraphics[width=0.535\textwidth]{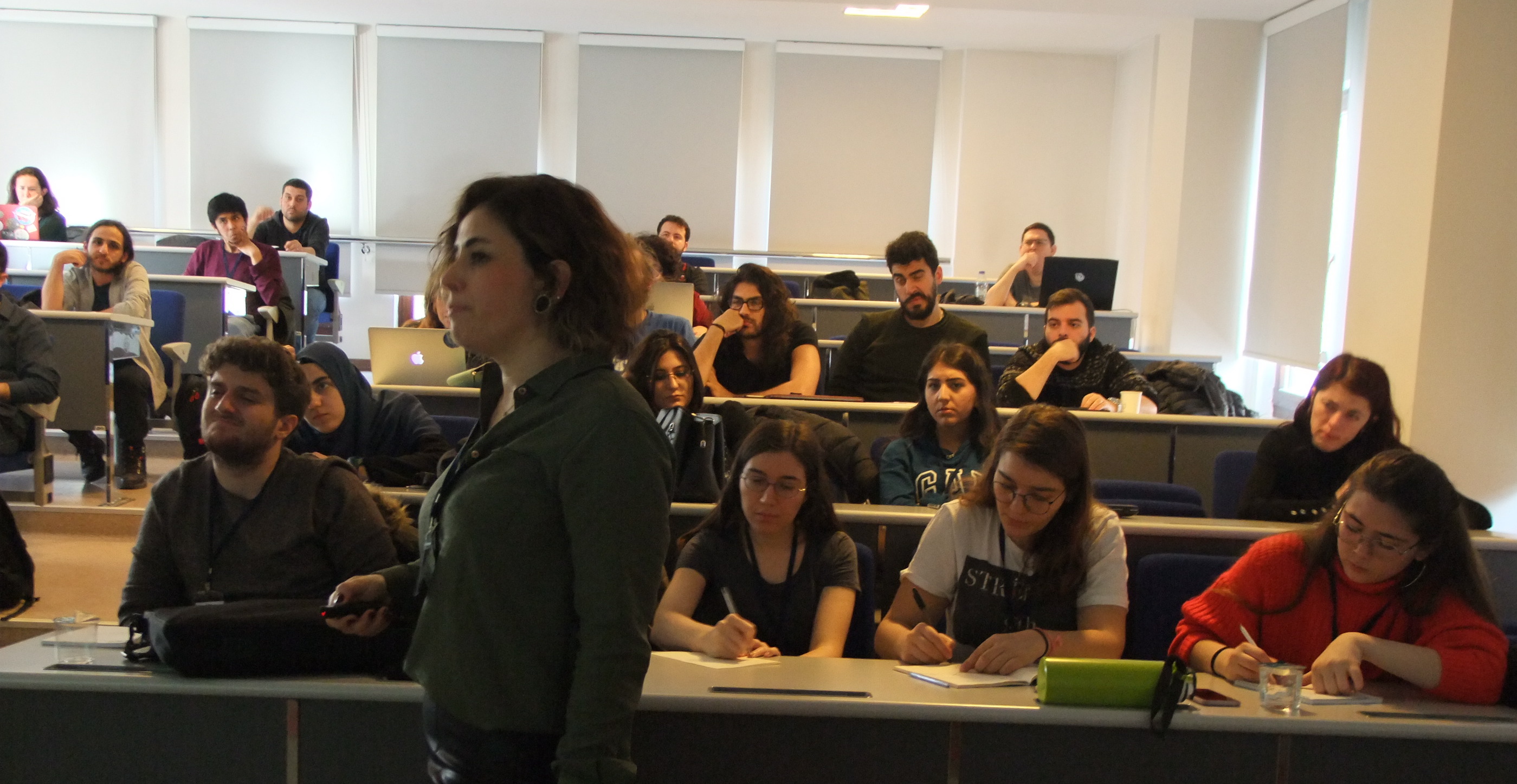}
    \includegraphics[width=0.365\textwidth]{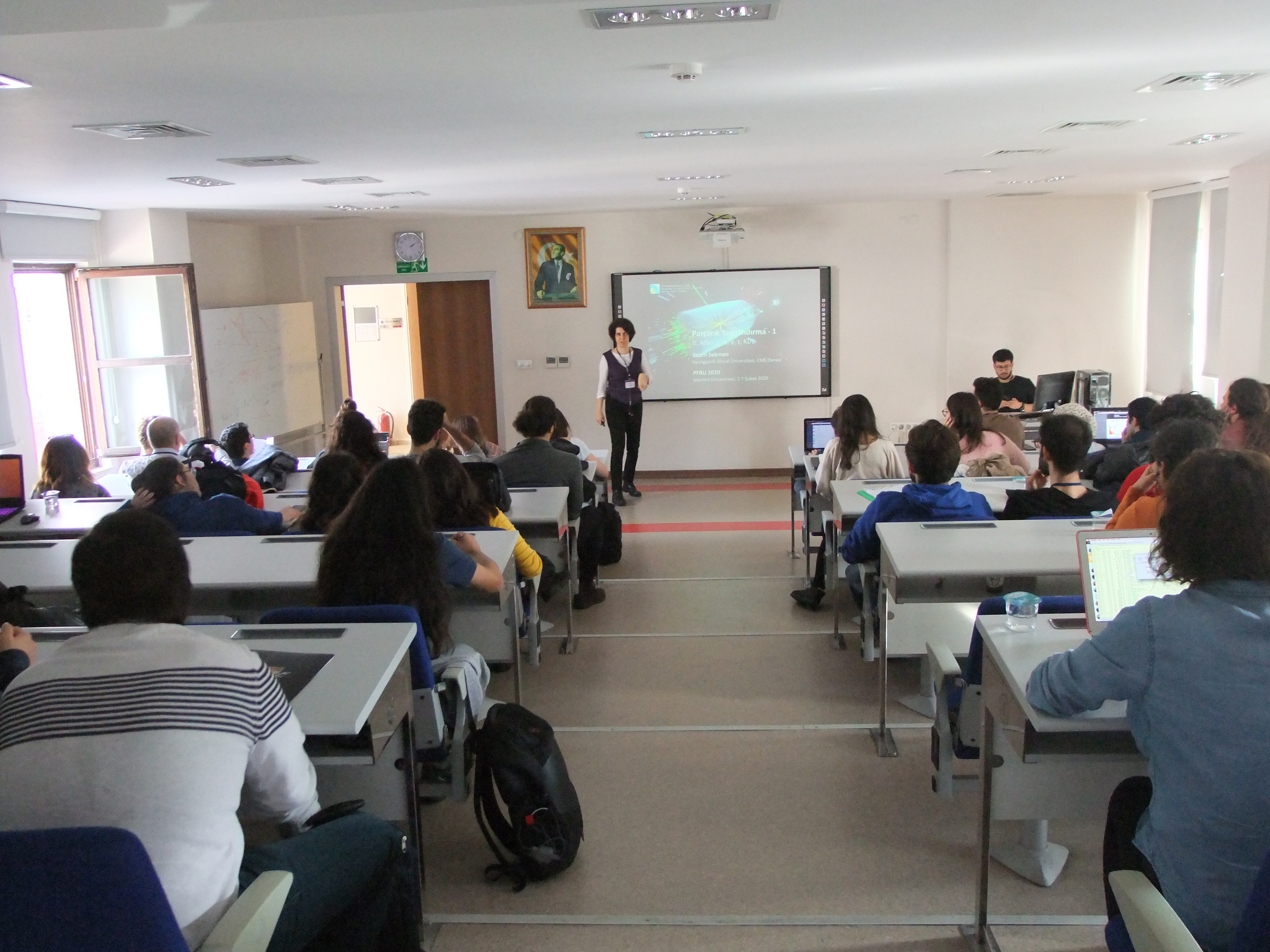}
    \includegraphics[width=0.45\textwidth]{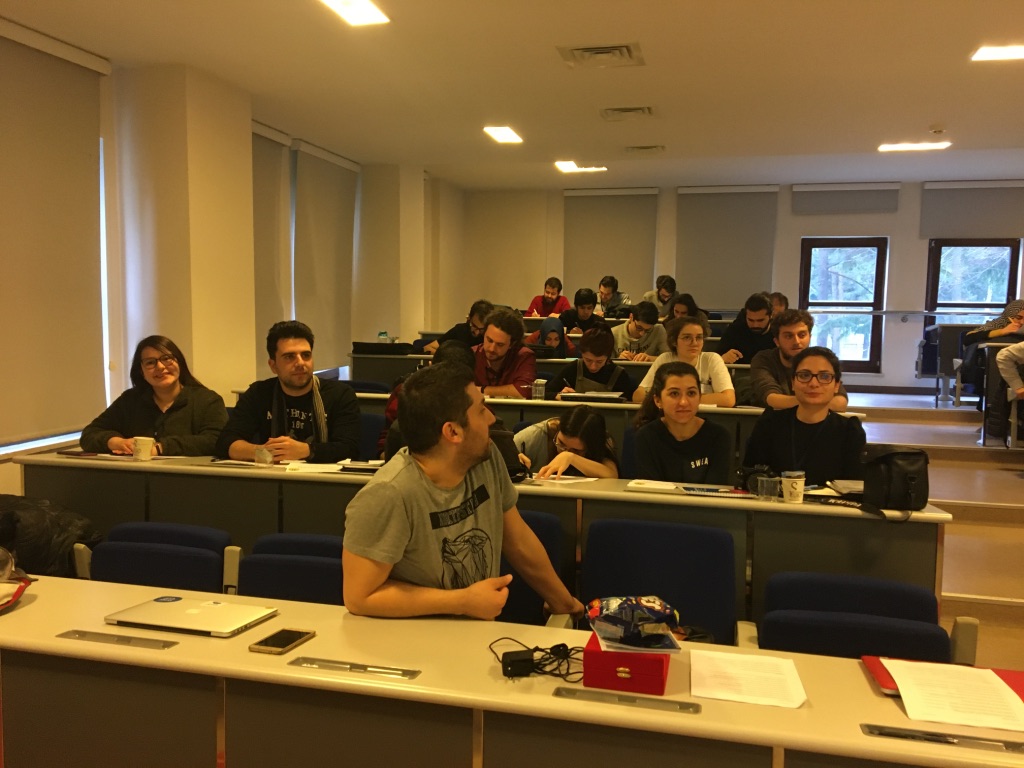}
    \includegraphics[width=0.45\textwidth]{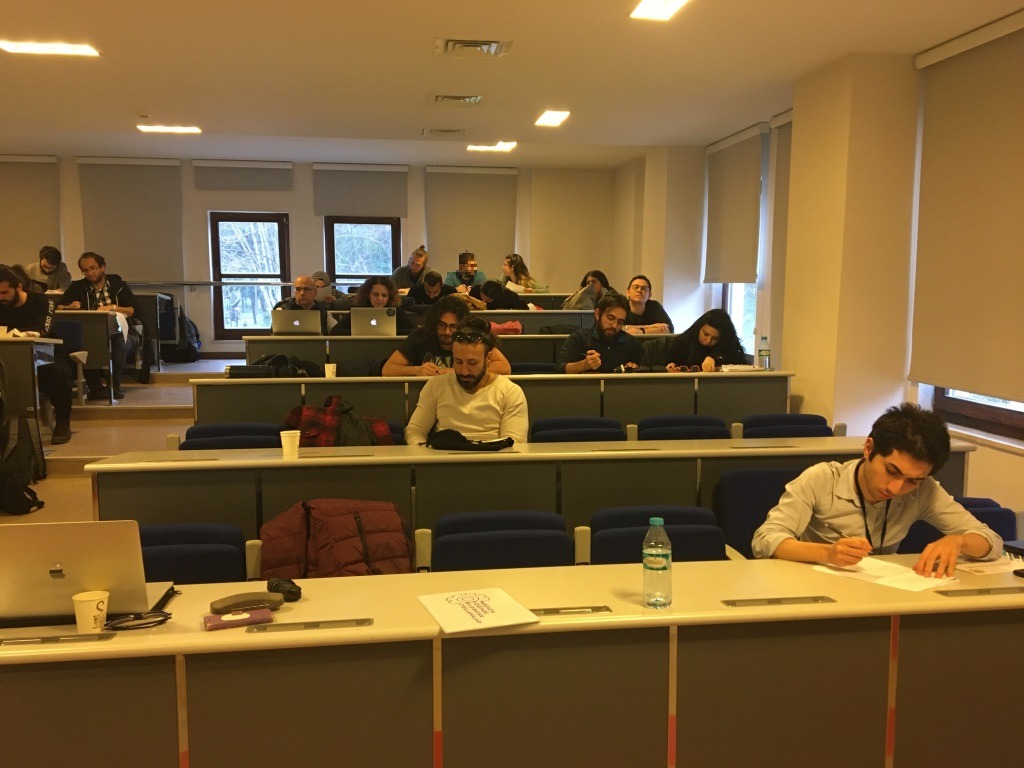}
    \caption{Snapshots from the lectures}
    \label{fig:lecture}
\end{figure}

\begin{figure}
    \centering
    \includegraphics[width=0.5\textwidth]{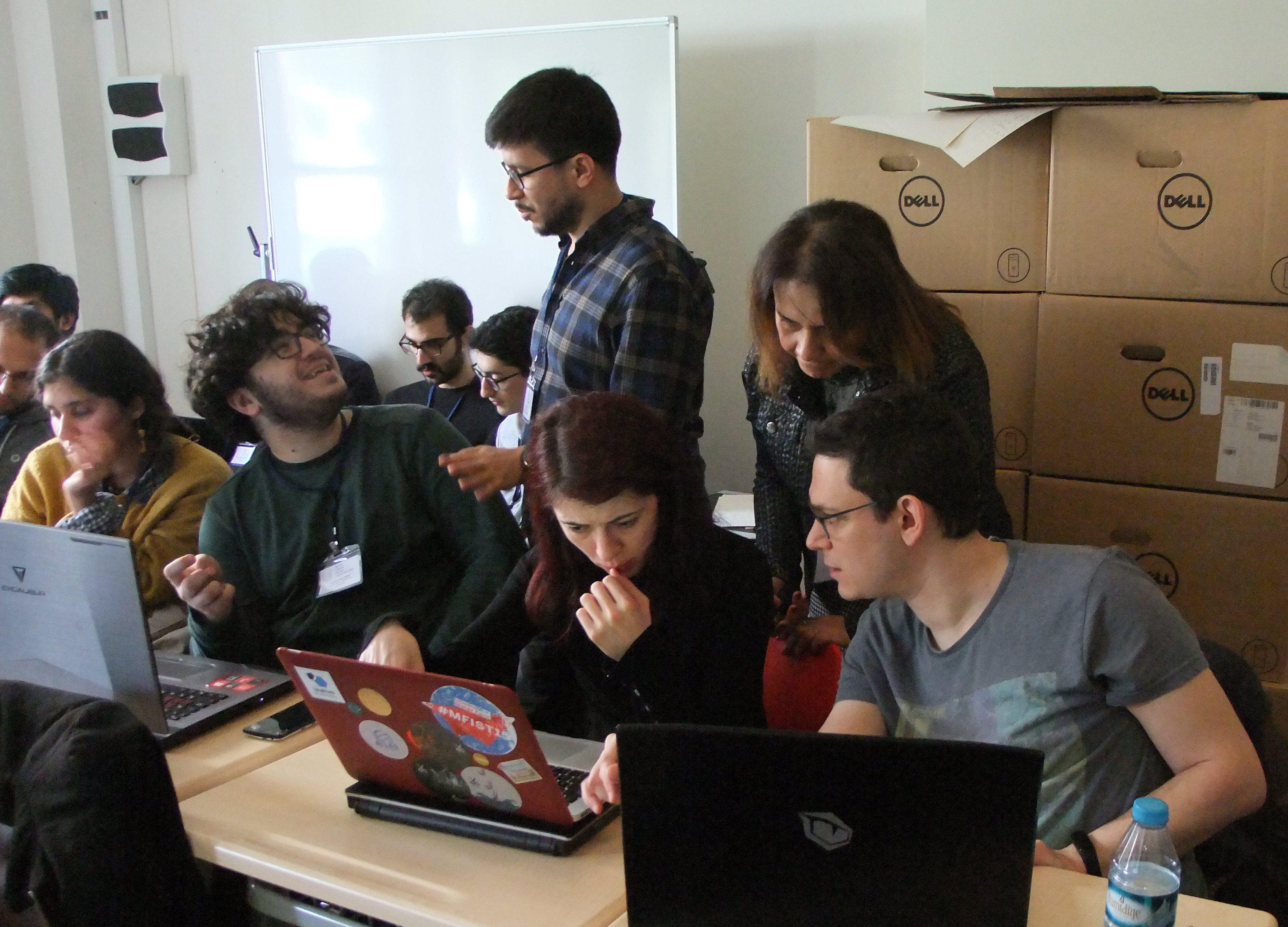}
    \caption{A snapshot from an exercise session }
    \label{fig:exercise}
\end{figure}

\begin{figure}
    \centering
    \includegraphics[width=0.4\textwidth]{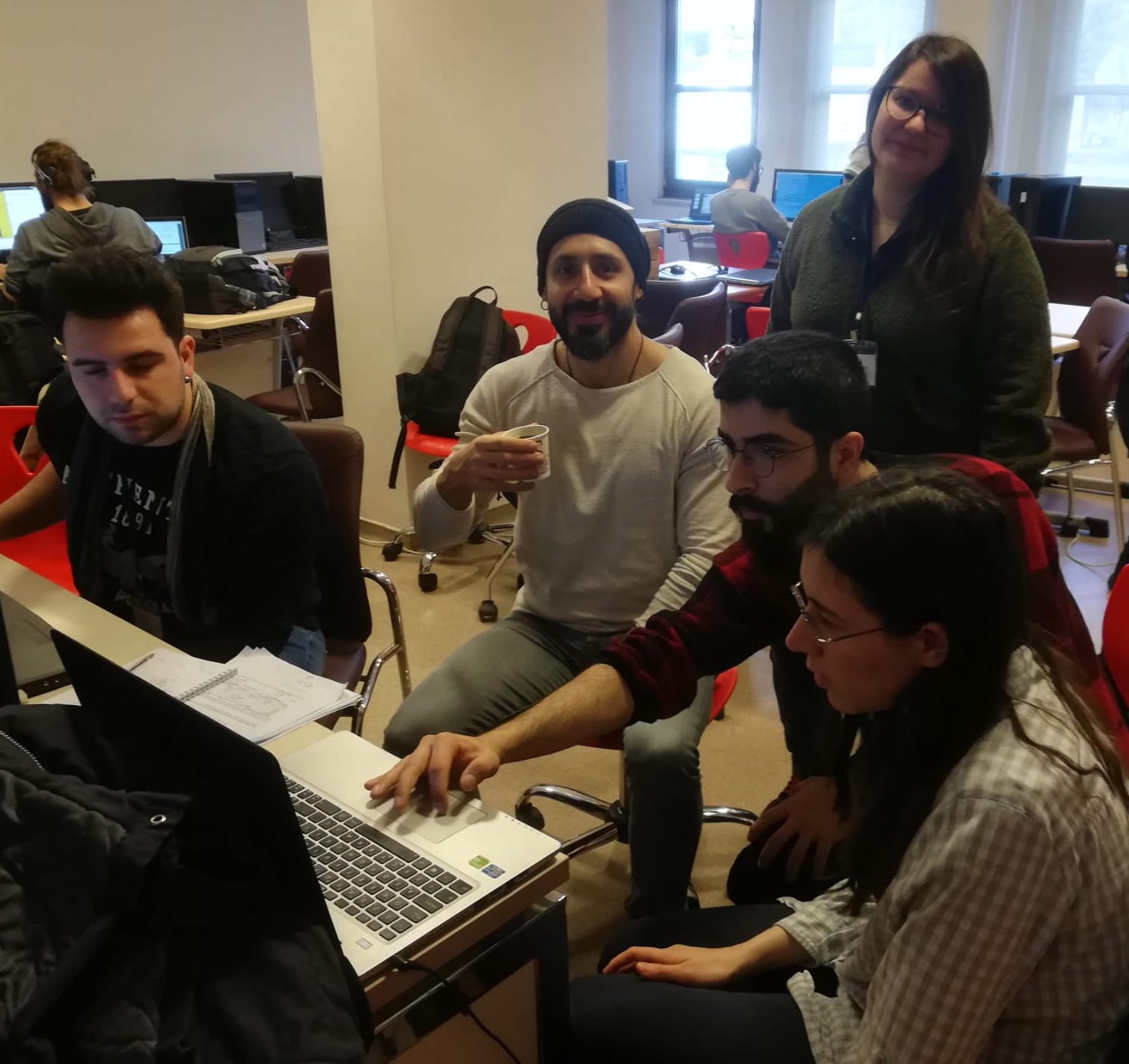}
    \includegraphics[width=0.5\textwidth]{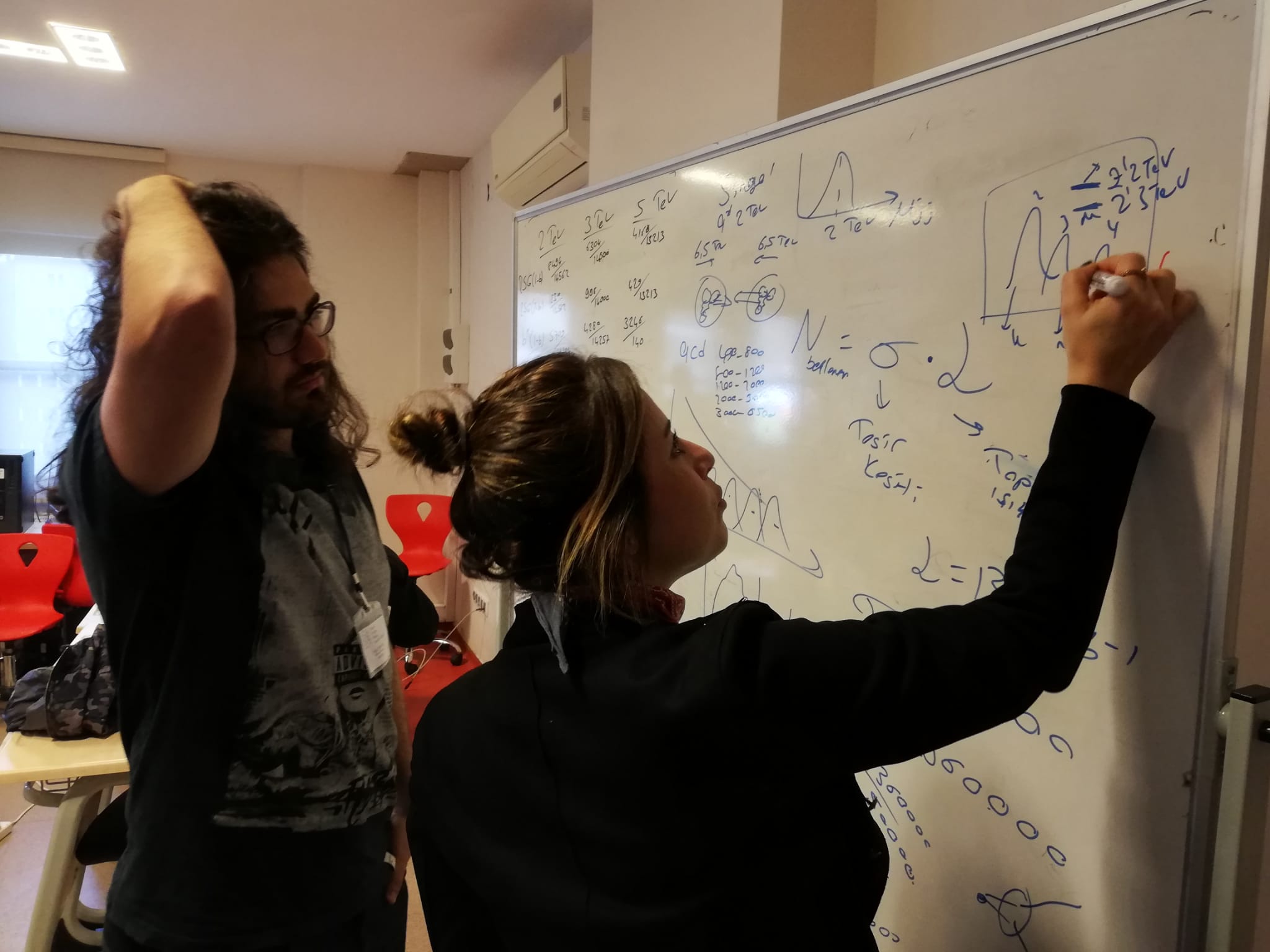}
    \caption{Snapshots from long exercises }
    \label{fig:longexercise}
\end{figure}

\begin{figure}
    \centering
    \includegraphics[width=0.5\textwidth]{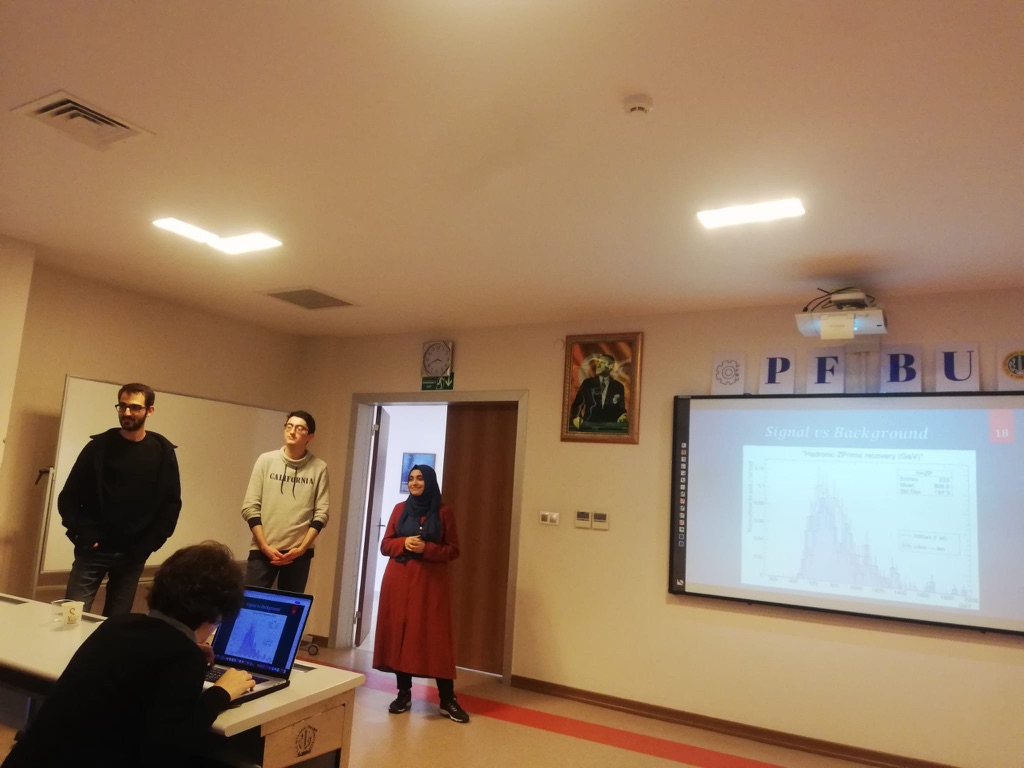}
    \caption{A snapshot from a group presentation}
    \label{fig:presentations}
\end{figure}

\begin{figure}
    \centering
    \includegraphics[width=0.5\textwidth]{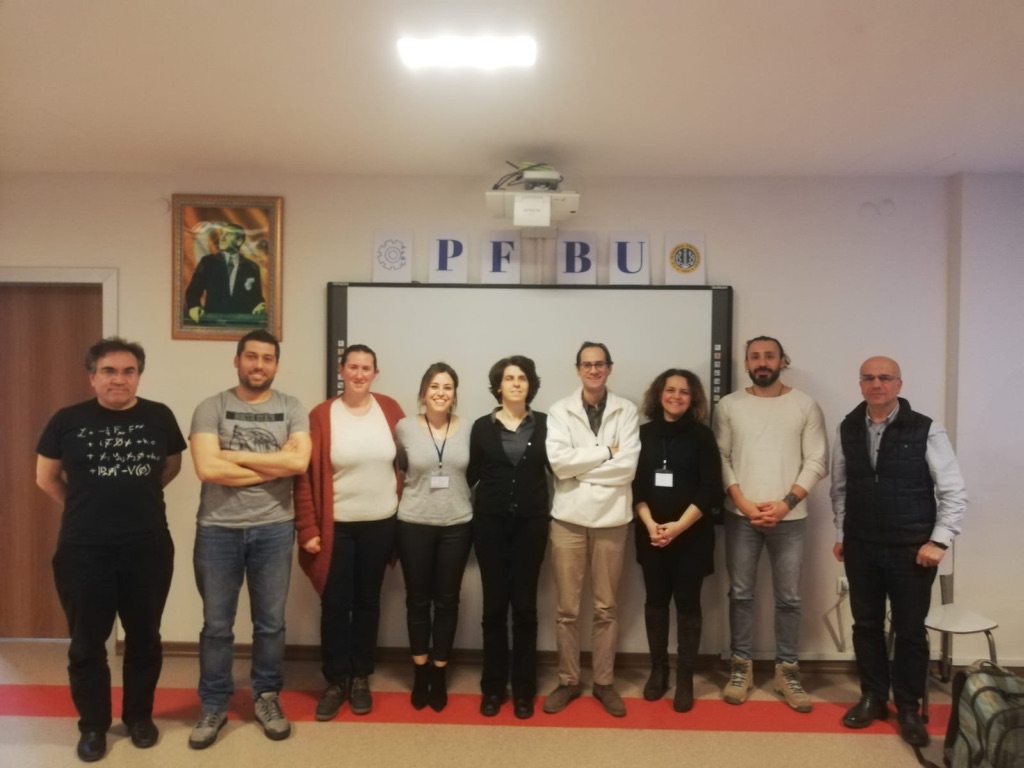}
    \caption{PFBU-5 Instructors and staff, including the authors of this manuscript.}
    \label{fig:instructors}
\end{figure}

\begin{figure}
    \centering
    \includegraphics[width=0.4\textwidth]{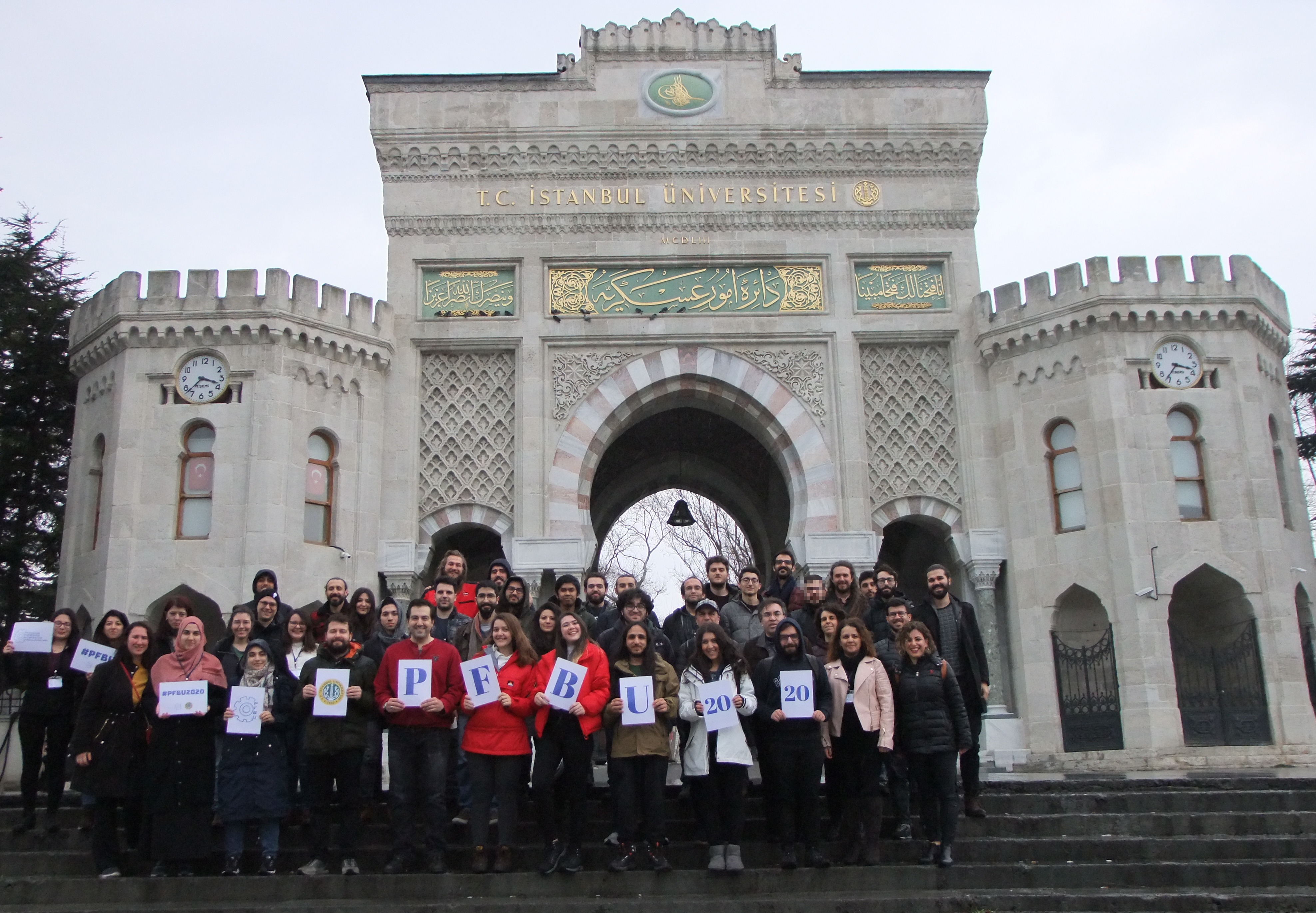}
    \includegraphics[width=0.5\textwidth]{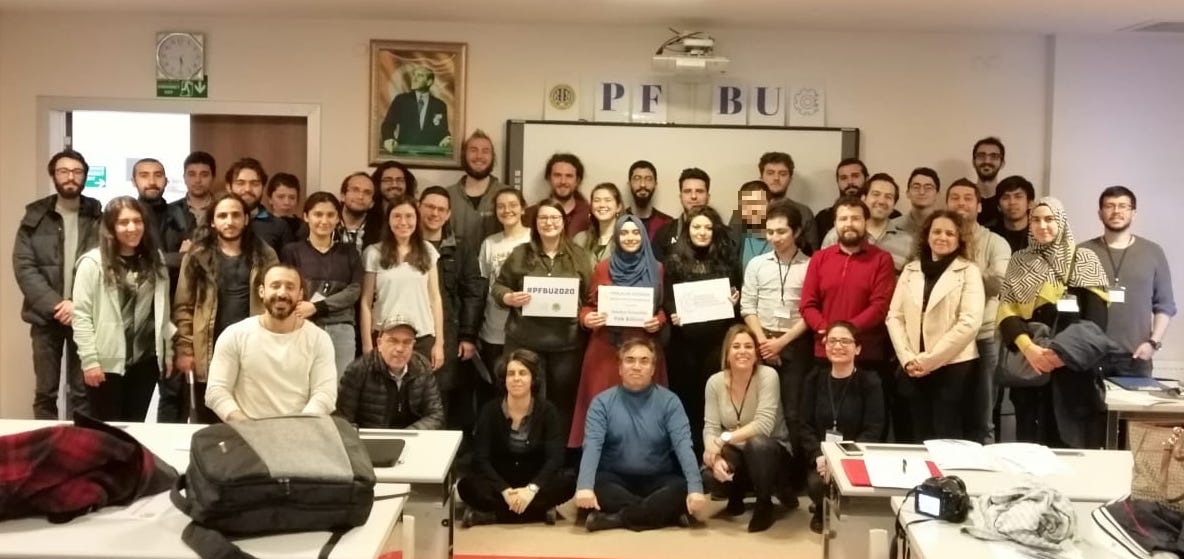}
    \caption{PFBU-5 Participants}
    \label{fig:participants}
\end{figure}

\newpage
\section{Further Details of the Short Exercises}
\label{appendix:B}

Here we give a complete description of the short examples serving to introduce CutLang, which were initially introduced in Section~\ref{sec:shortex}.

The first example is the Drell-Yan process of $pp \rightarrow Z \rightarrow e^{+} e^{-}$. The final state is two opposite-sign ``electrons''. First, events are required to have at least two electron candidates, and then the requirement for opposite-sign electric charge is applied. After each selection, a $Z$ boson candidate is reconstructed from the two highest $p_T$ electrons, and its mass is plotted. The CutLang ADL file for this example is:
\begin{lstlisting}
region   test 
  select ALL
  select Size(ELE) >= 2
  histo h1mReco,  "Z candidate mass (GeV)", 100, 0, 200, {ELE[0] ELE[1]}m
  select {ELE[0] ELE[1]}q==0
  histo h2mReco,  "Z candidate mass (GeV)", 100, 0, 200, {ELE[0] ELE[1]}m
\end{lstlisting}

As the second example, the Drell-Yan process of $pp \rightarrow Z \rightarrow \mu^{+} \mu^{-}$ is considered. The events are required to have at least two muons with  opposite-sign electric charge. After each selection, the $Z$ candidate mass is plotted. The CutLang ADL file for this process is similar to the second example.

The third example is the combined Drell-Yan process of $pp \rightarrow Z \rightarrow \mu^{+} \mu^{-} / e^{+} e^{-}$ . Since the final state is two opposite sign same flavour muons or electrons, events are categorised into the $ee$ and $\mu\mu$ channels.  At least two leptons are required to exist in the final state before the OS-SF criteria. The ADL file for studying this process would be:

\begin{lstlisting}
region   chele 
  select Size(ELE) >= 2
  select {ELE_0 ELE_1}q == 0 
  histo h1mReco,  "Z candidate mass (GeV)", 100, 0, 200, {ELE_0 ELE_1}m
  
region   chmuo 
  select Size(MUO) >= 2
  select {MUO_0 MUO_1}q == 0
  histo h2mReco,  "Z candidate mass (GeV)", 100, 0, 200, {MUO_0 MUO_1}m
\end{lstlisting}

In the fourth example, the same Drell-Yan process of the previous example is elaborated to specify ``good'' leptons. The leptons are defined as the union of electrons and muons. The good leptons are defined based on the cuts on their transverse momenta ($p_T>25$\,GeV) and pseudorapidity ($|\eta|< 2.5$). The students are expected to use CutLang syntax to find the best lepton combination to reconstruct a $Z$-boson candidate and they plot its invariant mass to comment on the decay width and eventually production cross section.

As the fifth example, the $pp \rightarrow W \rightarrow \mu \nu$ process is studied. The neutrino in the final state is assumed to have been measured as the missing transverse energy (MET), which, combined with the charged lepton, is used to reconstruct the $W$-boson candidate. In this case, the Lorentz vector from MET, METLV, is defined as ($MET, MET_x, MET_y, 0 $). Afterwards, the $p_{T}$ of the muon and the $W$ candidate mass are plotted. The ADL file for this example is:

\begin{lstlisting}
define Wreco : MUO[0] METLV[0]

region   testW 
  select ALL
  select Size(MUO) >= 1
  histo h1muPt,  "pT of muon (GeV)", 100, 0, 200, Pt(MUO[0])
  histo h2mReco,  "W candidate mass (GeV)", 100, 0, 200, m(Wreco)
\end{lstlisting}

The sixth example, which was already described in Section~\ref{sec:shortex} is the process $pp \rightarrow WW \rightarrow \mu \nu jj$ . The final state consists of a muon, two jets and MET. Initially both leptonically and hadronically decaying $W$ bosons are defined. The event selection starts with a requirement of at least one muon and sufficiently large MET ($>20$\,GeV). The events are also required to have at least two jets. Finally, the $W$ boson candidate masses are obtained from the highest momentum physics objects and are plotted for both leptonic and hadronic cases separately. The ADL file for this example is:

\begin{lstlisting}
define WLreco : MUO[0] METLV[0]
define WHreco : JET[0] JET[1]

region   test 
  select ALL
  select Size(MUO) >= 1
  select MET > 20
  select Size(JET) >=2 
  histo h1mWL,  "W_L candidate mass (GeV)", 100, 0, 200, {WLreco}m
  histo h2mWH,  "W_H candidate mass (GeV)", 100, 0, 200, {WHreco}m
\end{lstlisting}

The last short example is the process $pp \rightarrow t\bar{t} \rightarrow WWbb \rightarrow jjjjjj$. The final state contains at least six jets. Events are required to have two hadronically reconstructed $W$ bosons and two jets that can be identified as $b$-jets. Here, first the $W$s are defined from jets and then the optimal reconstruction is determined using a $\chi^2$ algorithm based on the known mass of the $W$ boson. Next, the top quarks are defined by pairing the reconstructed $W$ bosons and jets. The best pairing is again found via a $\chi^2$ algorithm without using any $b$-tagging. After all definitions, at least six jets are required. The MET in the event is expected to be small ($<100$\,GeV).  Finally, the total $\chi^2$ is minimized ($\chi^2~=0$) to plot the masses of the reconstructed objects. The ADL file for this example is given below:

\begin{lstlisting}
define WH1 : JET[-1] JET[-1]
define WH2 : JET[-3] JET[-3]
define Wchi2 : (({WH1}m-80.4)/2.1)^2 + (({WH2}m-80.4)/2.1)^2
define Top1 : WH1 JET[-2]
define Top2 : WH2 JET[-4]
define mTop1 : {Top1}m
define mTop2 : {Top2}m
define topchi2 : ((mTop1 - mTop2)/4.2)^2

region   best 
  select ALL
  select Size(JET) >= 6
  select MET < 100
  select Wchi2 + topchi2 ~=0
  histo hmWH1,  "Hadronic W reco (GeV)", 50, 50, 150, {WH1}m
  histo hmWH2,  "Hadronic W reco (GeV)", 50, 50, 150, {WH2}m
  histo hmTop1, "Hadronic Top reco (GeV)", 70, 0, 700, {Top1}m
  histo hmTop2, "Hadronic Top reco (GeV)", 70, 0, 700, {Top2}m
\end{lstlisting}


\begin{thebibliography}{9}

\bibitem {hpfbu} HPFBU Schools Main Webpage, \url{http://hpfbu.web.cern.ch/HPFBU/Anasayfa.html}.

\bibitem{Brooijmans:2016vro}
G.~Brooijmans, {\it et. al.}
``Les Houches 2015: Physics at TeV colliders - new physics working group report'', 
[arXiv:1605.02684 [hep-ph]].

\bibitem{Sekmen:2020vph}
S.~Sekmen, P.~Gras, L.~Gray, B.~Krikler, J.~Pivarski, H.~B.~Prosper, A.~Rizzi, G.~Unel and G.~Watts,
``Analysis Description Languages for the LHC'', 
PoS \textbf{LHCP2020} (2020), 065
[arXiv:2011.01950 [hep-ph]].

\bibitem{adlweb} Analysis Description Language Web Portal, \url{https://cern.ch/adl}.

\bibitem{Sekmen:2018ehb}
S.~Sekmen and G.~\"Unel,
``CutLang: A Particle Physics Analysis Description Language and Runtime Interpreter'', 
Comput. Phys. Commun. \textbf{233} (2018), 215-236
doi:10.1016/j.cpc.2018.06.023
[arXiv:1801.05727 [hep-ph]].

\bibitem{Unel:2019reo}
G.~Unel, S.~Sekmen and A.~M.~Toon,
``CutLang: a cut-based HEP analysis description language and runtime interpreter'', 
J. Phys. Conf. Ser. \textbf{1525} (2020) no.1, 012025
doi:10.1088/1742-6596/1525/1/012025
[arXiv:1909.10621 [hep-ph]].

\bibitem {cutlanggithub} CutLang Runtime Interpreter GitHub repository, \url{https://github.com/unelg/CutLang}.

\bibitem {RMT2} Barr A. et al, 2003, ``A variable for measuring masses at hadron colliders when missing energy is expected; $m_{T2}$: the truth behind the glamour'', J. Phys. G: Nucl. Part. Phys. 29 2343.

\bibitem {pfbuindico} PFBU 2020 School indico page,  \url{https://indico.cern.ch/event/877623/timetable/?view=standard}.

\bibitem {virtualbox} Virtual Box, \url{https://www.virtualbox.org/}.

\bibitem {root} ROOT Data Analysis Framework, \url{https://root.cern/}.

\bibitem {lexyacc} The Lex and Yacc Page, \url{http://dinasour.compilertools.net}.

\bibitem {opendata} CERN Open Data Portal, \url{http://opendata.cern.ch}.

\bibitem {pythia8} 
T.~Sjostrand, S.~Mrenna and P.~Z.~Skands,
``A Brief Introduction to PYTHIA 8.1'', 
Comput. Phys. Commun. \textbf{178} (2008), 852-867
doi:10.1016/j.cpc.2008.01.036
[arXiv:0710.3820 [hep-ph]].


\bibitem {slha} 
P.~Z.~Skands, B.~C.~Allanach, H.~Baer, C.~Balazs, G.~Belanger, F.~Boudjema, A.~Djouadi, R.~Godbole, J.~Guasch and S.~Heinemeyer, \textit{et al.}
``SUSY Les Houches accord: Interfacing SUSY spectrum calculators, decay packages, and event generators'', 
JHEP \textbf{07} (2004), 036.
doi:10.1088/1126-6708/2004/07/036
[arXiv:hep-ph/0311123 [hep-ph]].

\bibitem{hepmc} M.~Dobbs and J.~B.~Hansen,
``The HepMC C++ Monte Carlo event record for High Energy Physics'', Comput. Phys. Commun. \textbf{134} (2001), 41-46, doi:10.1016/S0010-4655(00)00189-2 .

\bibitem {delphes} 
J.~de Favereau \textit{et al.} [DELPHES 3],
``DELPHES 3, A modular framework for fast simulation of a generic collider experiment,''
JHEP \textbf{02} (2014), 057
doi:10.1007/JHEP02(2014)057
[arXiv:1307.6346 [hep-ex]].

\bibitem {adllhcanalyses} ADL file GitHub repository for LHC analyses, \url{https://github.com/ADL4HEP/ADLLHCanalyses}

\bibitem {intmasterclass} Johansson K. E. et al, 2007, ``European particle physics masterclasses make students into scientists for a day'', Phys. Educ. 42 636.

\end{thebibliography}
\end{document}